\documentclass[10pt, a4paper, conference]{IEEEtran}
\ifCLASSINFOpdf
\else
\fi
\usepackage{url}
\usepackage{multirow}
\usepackage{amssymb}
\usepackage{amsmath}
\usepackage{amsthm}
\usepackage[utf8]{inputenc}
\usepackage[english]{babel}
\usepackage{geometry}
\usepackage{mathtools}	
\usepackage{graphicx}
\usepackage{xcolor}

\usepackage{mathtools}
\usepackage[bookmarks=false]{hyperref}
\hypersetup{
	colorlinks=true,
	linkcolor=blue,
	filecolor=magenta,      
	urlcolor=cyan,
}

\usepackage[linesnumbered,ruled,vlined]{algorithm2e}
\usepackage{float}

\usepackage{graphicx}
\usepackage{caption,subcaption}

\usepackage{color, colortbl}
\definecolor{Gray}{gray}{0.8}

\begin{document}
%
\title{Threshold-Based Heuristics \\ for Trust Inference in a Social Network}

%

\author{\IEEEauthorblockN{Bithika Pal\IEEEauthorrefmark{0}, Suman Banerjee\IEEEauthorrefmark{0}, Mamata Jenamani\IEEEauthorrefmark{0}}
	
	\IEEEauthorblockA{\IEEEauthorrefmark{0}
		Department of Industrial and Systems Engineering,\\ Indian Institute of Technology, Kharagpur, 721302, India,\\ Email: bithikapal@iitkgp.ac.in, suman@iitkgp.ac.in, mj@iem.iitkgp.ac.in}
}


%


\maketitle

\begin{abstract}
Trust among the users of a social network plays a pivotal role in item recommendation, particularly for the cold start users. Due to the sparse nature of these networks, trust information between any two users may not be always available. To infer the missing trust values, one well-known approach is path based trust estimation, which suggests a user to believe all of its neighbors in the network. In this context, we propose two threshold-based heuristics to overcome the limitation of computation for the path based trust inference. It uses the propagation phenomena of trust and decides a threshold value to select a subset of users for trust propagation. While the first heuristic creates the inferred network considering only the subset of users, the second one is able to preserve the density of the inferred network coming from all users selection. We implement the heuristics and analyze the inferred networks with two real-world datasets. We observe that the proposed threshold based heuristic can recover up to 70 \% of the paths with much less time compared to its deterministic counterpart. We also show that the heuristic based inferred trust is capable of preserving the recommendation accuracy.

\end{abstract}


%
\IEEEpeerreviewmaketitle

\section{Introduction}

\textit{Social Network} is an interconnected structure among a group of interacting entities. Now\mbox{-}a\mbox{-}days these social networks are used by E\mbox{-}Commerce houses for different applications such as \textit{recommendation} \cite{seo2017personalized}, \textit{target advertisement} \cite{fan2014power}, and \textit{viral marketing} \cite{ashley2015creative}. In each of these applications, a crucial factor is the trust among the users, which is a measurement of belief one has with the other. In real world, the social network is sparse in nature and the trust value between any two specified user may not always be available. Naturally, the question arises, given a social network how to infer these missing trust values? This question forms the main basis of this paper. 

\par There are exists two approaches for trust inference; 1) graph theoretic \cite{josang2007survey, jiang2016understanding}, and 2) machine learning based \cite{Leskovec:2010:PPN:1772690.1772756}. One important property of trust is its propagation through social ties \cite{josang2007survey, jiang2016understanding}. This follows transitive nature; i.e., if a person $A$ trusts a person $B$ and $B$ trusts another person $C$, then $A$ also trusts $C$. Now, the trust weight value decays as the connecting path length between two users increases \cite{massa2007trust, golbeck2005personalizing} and diminishes after certain propagation length \cite{yuan2011small, pal2017trust}. The trust between two users' gets stronger if the connecting path between them passes through the highly influential nodes where the influence is the global reputation of a user in the whole system \cite{pal2017trust}. This incorporation of users global reputation for maximizing the local inferred trust gives better recommendation accuracy \cite{pal2017trust}. However, the key issue in this context is while maximizing the inferred trust, it has to search all the possible connecting paths between two users, up to certain prespecified length. In this work, we propose threshold based heuristics to reduce this search space and get the prediction of inferred trust as close as possible to its deterministic counterpart. We use this inferred trust values for the recommendation in two real-world datasets. We show empirically that around 70\% of the trust can be recovered from its enumeration based counterpart. Our contributions in this paper are fourfold:
\begin{itemize}
	\item Two heuristics are proposed for path based trust inference.
	\item One new node threshold function is proposed based on both trust network and the rating data.
	\item Comparison of all the heuristics are done for different propagation length and different node threshold cut-off.
	\item Effect in recommendation accuracy is analyzed using the proposed heuristics.
\end{itemize}

The paper is organized as follows. In Sec. \ref{sec:relwork}, some of the key related works are mentioned from the literature. The proposed methodology and the experimental results are given in Sec. \ref{sec:proppsed} and \ref{sec:results}, respectively. Finally, Sec. \ref{sec:conclusion} concludes the work.
	
\section{Related Work} \label{sec:relwork}
This section briefs the literature in two parts, firstly, the related work in trust inference and, secondly, the review of trust-aware recommender systems. First work in this direction is trust inference from explicit binary trust network considering its linear decay with the connecting path length \cite{massa2004trust}. In \cite{massa2007trust}, the authors proposed a path-based inference method in weighted trust graph settings. This method does the average of trust weights coming from all its incoming path from source to destination (\textit{MoleTrust}). In \cite{golbeck2005personalizing}, the same concept is used for trust inference between two users with the constraint of path capacity to limit the trust propagation (\textit{TidalTrust}). In \cite{yuan2011small}, the authors mentioned the trust propagation length as the average path length of the network. The authors also empirically showed that beyond this length, trust propagation does not lead to any further improvement in recommendation accuracy. In \cite{Moradi:2015:RRM:2827896.2828216}, the authors proposed an incremental method for trusted neighbor selection from the inferred trust network using their developed reliability metric to improve recommendation accuracy. All these studies inferred the local trust obtained from the connecting path. However, none of them considered the individual node influence in the trust inference. In \cite{pal2017trust}, the authors took into account this individual influence in local trust inference. The main intuition in their study is that `trust more a user if it is already trusted by a larger number of users'. There is also algebraic methods of trust inference where partial order relation of trust is in forced \cite{Gao:2016:SST:2959100.2959148}. In this paper, we propose threshold based heuristics for trust inference using individual node influence. We also design a metric to capture this node influence from trust network and rating data.

\par Now, Trust-aware or social recommender system (TARS) has also two types of settings for rating prediction problem. One is the use of inferred trust in memory based collaborative filtering where the trust is used in place of user-user similarity for rating prediction \cite{massa2007trust, golbeck2005personalizing, yuan2011small, pal2017trust, massa2004trust, Moradi:2015:RRM:2827896.2828216}. The other one is the incorporation of social or trust information in model-based collaborative filtering technique. These methods are based on the addition of trust or social relation as the regularization in the existing matrix factorization model that considers trust matrix having low rank. Based on the mapping techniques of latent features from the social users and rating information, different models are available in literature \cite{Jamali:2010:MFT:1864708.1864736, zhou2012kernelized, yang2017social, guo2016novel}. In this study, we use our inferred trust in the memory based TARS.

\section{Proposed Path Based Trust Inference Methodology} \label{sec:proppsed}

We consider the social network to be represented by a \textit{vertex} as well as \textit{edge weighted, directed graph} $\mathcal{G}(\mathcal{U}, \mathcal{E}, \mathcal{T}, \theta)$. Here, $\mathcal{U}(\mathcal{G})=\{u_1, u_2, \dots, u_{n}\}$ is the set of users, $\mathcal{E}(\mathcal{G})$ is the set of \textit{directed social ties} among the users, \textit{i.e.}, $\mathcal{E}(\mathcal{G})\subset \mathcal{U}(\mathcal{G}) \times \mathcal{U}(\mathcal{G})$. $\mathcal{T}$ is the edge weight function, which maps each edge to a fraction between $0$ and $1$; i.e.; $\mathcal{T}: \mathcal{E}(\mathcal{G}) \rightarrow [0,1]$. For a particular social tie $(u_iu_j) \in \mathcal{E}(\mathcal{G})$, we denote its \textit{tie strength} as $t_{ij}$, which signifies the amount of \textit{local trust}, the user $u_i$ has on $u_j$. $\theta$ is the vertex weight function, which maps each user of the network to a real value; i.e.; $\theta:\mathcal{U}(\mathcal{G}) \rightarrow \mathbb{R}^{+}_0$ \footnote{$\mathbb{R}^{+}_0$ denotes the set of all +ve real numbers including zero.}. For the node $u_i \in \mathcal{U}(\mathcal{G})$, its vertex weight (often called as \textit{reputation} or \textit{trustworthiness}) is denoted by $\theta_{i}$. It signifies the amount of \textit{influence} the user $u_i$ has in the whole system. The symbols and notations with their meanings are discussed in Table \ref{Tb:symb}.

\begin{table}[!htb]
	\centering
	\caption{\fontsize{8}{8}\selectfont{\uppercase{Description of the Notations}}} \label{Tb:symb}
	\begin{tabular}{|p{0.9cm}|p{7cm}|}
		\rowcolor{Gray}
		\hline
		Symbols & Definitions \\
		\hline
		$\mathcal{G}$, $\hat{\mathcal{G}}$ & Given and inferred social network \\
		\hline
		$\mathcal{U}, u_i, \theta_i$ & The set of users, the $i$-th user and the node weight of $u_i$ \\
		\hline
		$p_k, l_k$ & Path between any arbitrary two nodes and length of that path \\
		\hline
		$\mathcal{P}, \mathcal{P}(l_k)$ & Penalty function and its value for the length $l_k$\\
		\hline
		$\mathcal{B}$ & Benefit function \\
		\hline
		$ \mathcal{B}(\theta)$ & Benefit from a path with all of its intermediate nodes' weight \\
		\hline
		$L_{max}$ & Maximum allowable trust propagation length \\
		\hline
		$t_{ij}, \hat{t}_{ij}$ & Trust, inferred trust value from user $u_i$ to $u_j$ \\
		\hline
		$\Gamma_{\mathcal{G}}^{l}(u_i)$ & The neighbors of user $u_i$ in  $\mathcal{G}$ at distance $l$\\
		\hline
		$\theta^c$ & Cut-off threshold to select a set of nodes for propagation \\
		\hline
		$\mathcal{X}$ & A subset of nodes from $\Gamma_{\mathcal{G}}^{l}(u_i)$ for propagation \\
		\hline
		$\mathcal{I}, i_j$ & The set of items, the $i$-th item \\
		\hline
		$r_{ij}, \hat{r}_{ij}$ & Rating, predicted rating user $u_i$ to item $i_j$ \\
		\hline
		$\bar{r_{i}}$ & Average rating of user $u_i$ \\
		\hline
			
	\end{tabular}
\end{table}
 
\par As mentioned previously, real life social networks are extremely sparse. Hence, local trust value between two arbitary users may not be available. However, it can be computed using some inference mechanism. Now, we formally state the trust inference problem from the literature \cite{pal2017trust}. Suppose, for two arbitary users $u_i,u_j \in \mathcal{U}(\mathcal{G})$, their local trust $t_{ij}$ needs to be inferred. Assume, $u_i$ and $u_j$, are connected by $K$ number of simple paths $p_1,p_2,\dots, p_k, \dots p_K$ (each path $p_k \equiv \langle u_i,u_1,\dots, u_x, \dots u_{l_{k-1}} u_j \rangle$) with the corresponding length $l_1,l_2,\dots, l_k, \dots, l_K$, where $2 \leq l_k \leq L_{max}(\text{maximimum allowable propagation length})$. Now, the local trust inferred from each path is associated with two factors: 1) the penalty $\mathcal{P}(l_k)$, an increasing function on the path length and 2) the benefit $\mathcal{B}(\theta)$ from all the intermediate nodes in that path. We want that the inferred trust between any two users is as large as possible among $K$ paths. It leads to the following mathematical formulation of the problem (Equation \ref{Eq:Formulation}) where $l_k=length(p_k)$.

\begin{equation} \label{Eq:Formulation}
t_{i,j}=\underset{\substack{{p_k \in \{p_1,\dots p_K \}}} }{argmax} \left \{1-\mathcal{P}(l_k)+\sum_{\substack{x=1 \\ u_x \in p_k}}^{{l_{k-1}}} \mathcal{B}(\theta_x) \right \}
\end{equation}

$\mathcal{P}(l_k)$ is the penalty incurred for taking path $p_k$ and $1-\mathcal{P}(l_k)$ represents the gain for choosing the path $p_k$. Also, $\sum \mathcal{B}(\theta_x)$ adds more value on $t_{ij}$ to rank the neighbors of $u_i$ at length $l_k$. Thus, Equation \ref{Eq:Formulation} captures the contributions in the inferred trust between any two nodes from both, the path length and the individual node influence in that path.


Considering linear decay along the path, $\mathcal{P}(l_k)$ is chosen as $(l_k-1)/L_{max}$ \cite{ yuan2011small, pal2017trust, massa2004trust, Moradi:2015:RRM:2827896.2828216}. $L_{max}$ is chosen as the average path length of the network in the previous studies \cite{yuan2011small,Moradi:2015:RRM:2827896.2828216, pal2017trust}. The concept of $\mathcal{B}(\theta)$ is introduced by the authors of \cite{pal2017trust}, which enforces to search all the possible paths between two users. In absence of $\mathcal{B}(\theta)$, $t_{ij}$ comes from the shortest path \cite{ yuan2011small, massa2004trust,   Moradi:2015:RRM:2827896.2828216}. Incorporating influence of the intermediate nodes and sometimes going beyond the shortest path actually helps in improving the recommendation accuracy \cite{pal2017trust}. Now, in this problem setting the main drawback is its scalability. As the path lengths increase, an exponential growth of the problem complexity is observed. To overcome this situation, we introduce two threshold based heuristic methods to reduce the search space of the number of paths.

\subsection{Heuristic Based on Threshold Cut-Off} \label{sec:threshold_heuristic}
In searching all possible paths between the nodes $u_i$ and $u_j$, one trivial approach is to start from the node $u_i$ and find its $\Gamma_{\mathcal{G}}^{1}(u_i)$, $\Gamma_{\mathcal{G}}^{2}(u_i)$, $\dots$, $\Gamma_{\mathcal{G}}^{L_{max}}(u_i)$, where $\Gamma_{\mathcal{G}}^{l}(u_i)$ denotes the neighbors at distance $l$ from $u_i$. Here, in every propagation, the direct neighbors of $\Gamma_{\mathcal{G}}^{l}(u_i)$ build $\Gamma_{\mathcal{G}}^{l+1}(u_i)$. Hence, the purpose of the heuristic is to get a subset $\mathcal{X}$ of $\Gamma_{\mathcal{G}}^{l}(u_i)$ at every propagation, which is to be used in building $\Gamma_{\mathcal{G}}^{l+1}(u_i)$. As $\vert \mathcal{X} \vert < \vert \Gamma_{\mathcal{G}}^{l}(u_i) \vert$, this will definitely reduce the search space. Now, the question arises how to get these subsets in every propagation? As the intuition is to get the maximum $t_{ij}$, it has to pass through more influential nodes (higer $\theta$ value), so that, the contribution of $\mathcal{B}(\theta)$ becomes higher. For this purpose, we define a cut-off threshold value $\theta^c$ at every propagation starting from a node $u_i$, and the nodes having higher $\theta$ than the cut-off constitutes the subset for the next propagation. Mathematically, the subset $\mathcal{X}$ at the propagation length $l$ starting from $u_i$ is $ \{w: \theta_w \geq \theta^c, \forall w \in  \Gamma_{\mathcal{G}}^{l}(u_i)\}$. As, at every propagation it greedily chooses the higher $\theta$ valued nodes, the inferred trust from the heuristic $t_{ij}^h$ will be closer to the optimal $t_{ij}$ from Equation \ref{Eq:Formulation}. In this heuristic, the value of $\theta^c$ at every step is computed as the average of the nodes weights in the current nodes immediate neighborhood, given in Equation \ref{Eq:theta_c}.

\begin{equation} \label{Eq:theta_c}
\theta^c = \frac{c_{th} \left (\sum_{w \in \Gamma_{\mathcal{G}}^{l}(u_i)}  \theta_w \right)}  { \vert \Gamma_{\mathcal{G}}^{l}(u_i) \vert}
\end{equation}

Here, $c_{th}$ is a constant to scale the cut-off threshold value.

\subsection{Different Types of Vertex Weight Function }
Based on different intuition, several vertex weight functions can be designed in this regard. Here, we present three ways to set the $\theta$ value for a node.
\subsubsection{Indegree as Weight} \label{sec:deg_weight}
The intuition here is if a user is trusted by many users, it implicitly signifies the user has higher influence or reputation in the whole system. Hence, user can be considered as more trustworthy. Therefore, we propose that the in-degree of a vertex can be used as the vertex weight $\theta$, i.e., $\theta_i := indeg(u_i)$. This brings down our problem into a more general setting of finding a path between two nodes of certain length passing through the high degree vertices.  

\subsubsection{Degree-of-Trustworthiness as Weight} \label{sec:delta_weight}
The intuition mentioned in Sec. \ref{sec:deg_weight} is captured in a metric named \textit{degree-of-trustworthiness} ($\delta_i$) \cite{pal2017trust}. The authors define $\delta_i$ as Equation \ref{eq:delta}. Here, we propose to use the weight $\theta_i$ for a user node $u_i$ as degree-of-trustworthiness, $\delta_i$.

\begin{equation} \label{eq:delta}
\delta_i =  \frac{q  \cdot indeg(u_i)}{  (max(indeg(\mathcal{G})) + \epsilon)} ,\  \epsilon \geq zero, \  0 < q \leq {1}/{L_{max}}
\end{equation}

\subsubsection{Degree-of-TrustNPurchase as Weight} \label{sec:gamma_weight}
Here, we present a new vertex weight function influenced by both degree-of-trustworthiness and users purchase pattern. Along with trusted by many other users, if a user purchases more items, believing in that user is more helpful. Now, in purchase pattern, we try to capture users' significance based on the different category of items selection. We divide the whole item set $\mathcal{I}$ into three categories, heavily rated items $\mathcal{I}^H$, average rated items $\mathcal{I}^A$, and cold start items $\mathcal{I}^C$. The set of heavily rated, average rated and cold start items purchased by the user $u_i$ is denoted by $\mathcal{I}^H_{u_i}$, $\mathcal{I}^A_{u_i}$, and $\mathcal{I}^C_{u_i}$ respectively. Now, the \textit{Degree-of-TrustNPurchase} for the user $u_i$ and symbolized by $\gamma_i$, can be calculated as Equation \ref{eq:gamma} and $\theta_i := \gamma_i$.

\begin{equation} \label{eq:gamma}
\begin{aligned}
\gamma_i =   \frac{\vert \mathcal{I}^H_{u_i} \vert}{max_{u_i \in \mathcal{U}} \ \mathcal{I}^H_{u_i}}
+ \frac{\vert \mathcal{I}^A_{u_i} \vert}{max_{u_i \in \mathcal{U}} \ \mathcal{I}^A_{u_i}}
+ \frac{\vert \mathcal{I}^C_{u_i} \vert}{max_{u_i \in \mathcal{U}} \ \mathcal{I}^C_{u_i}}\\
	 + \frac{indeg(u_i)}{max(indeg(\mathcal{G})) } \quad \quad \quad \quad 
\end{aligned}
\end{equation}

The individual item category plays a pivotal role. Like, heavy rated items are less in the count, so the ratio is more likely to be greater. This captures it's more likeliness to other users. Whereas, cold rated items are more in number, so it is very likely to have less value in that ratio. However, large ratio value from cold item part signifies that the user has potential to be more trusted.

\par Now, we discuss about $\mathcal{B}(\theta)$ from the vertex weight of the intermediate nodes. For the weight functions mentioned in Sec. \ref{sec:deg_weight}, \ref{sec:delta_weight}, $\mathcal{B}(\theta_i)$ is used as the \textit{degree-of-trustworthiness} ($\delta_i$). For the \textit{Degree-of-TrustNPurchase} in Sec. \ref{sec:gamma_weight}, $\mathcal{B}(\theta) = \sigma(\sum_x \gamma_x)/L_{max}$, where $u_x$ is an intermediate node of the connecting path. Sigmoid ($\sigma$) function is used to bring the value in [0,1].

\subsection{Algorithm For Threshold Based Heuristic} \label{sec:algo_threshold_heuristic}
In this section, Algorithm \ref{alg:1} describes the procedure for finding the paths between any two nodes in the social network $\mathcal{G}$ using the threshold based heuristic (mentioned in Sec. \ref{sec:threshold_heuristic}).
\begin{algorithm}[h]
	\caption{Algorithm for calculating heuristic based paths up to length $L_{max}$ in social network $\mathcal{G}$}
	\label{alg:1}
	\KwData{Social Network $\mathcal{G}$, integer $L_{max}$\\}
	\KwResult{List of paths indexed with vertex pair, $hPaths$ }
	
	\SetKwFunction{Fneighboradd}{addNeighbor}
	
	$\textbf{global } hPaths=NULL$\;

	\SetKwProg{Pn}{Function}{:}{\KwRet}
	\Pn{\Fneighboradd{$sNode$, $cNode$, $\mathcal{G}$, $L$, $path$}}{
		
		\If{ $L==1$}{  
			\textbf{return}\;
		}
		Calculate $\theta^c$ from $\mathcal{G}.succesors(cNode)$  \tcp{Eq. \ref{Eq:theta_c}}
		\For{$i \in \mathcal{G}.succesors(cNode)$}{
			\If{$i \neq sNode$ and $i \notin path$}{ 
				$tempPath = path $\;
				$tempPath.append(i)$\;
				\If{ $! \ \mathcal{G}.hasEdge(sNode,i)$}{  
					$hPaths[\langle sNode,i \rangle].append(tempPath)$\;
				}
				\If{ $c_{th}.\theta_i \geq \theta^c$}{  
					\texttt{addNeighbor}{($sNode$, $i$, $\mathcal{G}$, $L-1$, $tempPath$)}\;
				}
			}	
		} 
	}
	
	\For{each node $ i \in \mathcal{G}.nodes()$ }{ 
		\For{$ j \in \mathcal{G}.succesors(i)$ }{ 
			$path=[i,j]$ \tcp*{List of nodes}
			\texttt{addNeighbor}{($i$, $j$, $\mathcal{G}$, $L_{max}$, $path$)}\;
		}
	}	
\end{algorithm}

\begin{figure}
	\centering
	\includegraphics[scale=0.7]{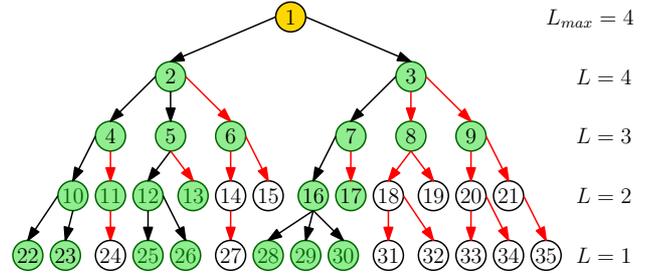}
	\caption{  \fontsize{8}{10}\selectfont{ Example for a sequence of propagation in the threshold based heuristic. Starting from node $1$, all the green colored nodes are reached using heuristic of Algorithm \ref{alg:1} and added into $hPaths$. The red marked arrows signify to neglect the nodes for further propagation. In the diagram, all the nodes are presented with their decreasing vertex weight $\theta$, from left to right. } } \label{fig:example}
\end{figure}

 For each node in the network $\mathcal{G}$, the recursive function \texttt{addNeighbor()} is called for all of its immediate succesors or neighbors. The function \texttt{addNeighbor()} has five input parameters, the starting node ($sNode$), the current node from which direct neigbors has to be found ($cNode$), the network $\mathcal{G}$, distance to propagate further ($L$), list of nodes in the path from $sNode$ to $cNode$ ($path$). In each function call, it  computes the cut of threshold from $cNode$; adds each immediate neighbors of $cNode$ to the $path$ and makes an entry to the global dictionary $hPaths$ with index $\langle SNode, neighbor \rangle$; finally, calls the recursive function with $L-1$ for all the nodes in the subset $\mathcal{X}$ formed by the cut-off thresold $\theta_c$. This process continues till $L$ reaches to 1. The time complexity for each node can be given as $T(L)=dT(L-1)+ d \mathcal{O}(L) + d$, if we consider a regular graph with degree $d$. This is reduced to $\mathcal{O}(L_{max}d^{L_{max}})$. For all the nodes, the complextiy is $\mathcal{O}(nL_{max}d^{L_{max}})$. As every recursion of Algorithm \ref{alg:1} always propagate for less than $d$ nodes, the time complexity of the algorithm becomes $o(nL_{max}d^{L_{max}})$. Now, as the social network follows power law degree distribution, $d$ can be replaced with the expected degree of a node, which is $\sum_{k=1}^{\Delta} k.c.k^{-\gamma}$ ($\Delta$ is the maximum degree of the network and c is constant). One example propagation is shown in Figure \ref{fig:example}.

\begin{algorithm}[h]
	\caption{Algorithm for creating inferred network $\hat{\mathcal{G}}$}
	\label{alg:2}
	\KwData{Social Network $\mathcal{G}$, $hPaths$\\}
	\KwResult{Inferred social network $\hat{\mathcal{G}}$}
	
	$\hat{\mathcal{G}} \leftarrow \mathcal{G} $\;
	
	\For{each vertex pair $ \langle i,j \rangle \in hPaths.indicies()$ }{ 
		$pathList=hPaths[\langle i,j \rangle]$\;
		$t_{ij} = 0$\;
		\For{ path $ k \in pathList$ }{ 
			
			Calculate $\mathcal{P}(len(k))$\;
			Calculate $\mathcal{B}(\theta)$ from path $k$\;
			$\hat{t}_{ij} = 1 - \mathcal{P}(len(k)) + \mathcal{B}(\theta)$\;
			\If{$\hat{t}_{ij} > t_{ij}$}{$ t_{ij}=\hat{t}_{ij}$\;}
		}
		$\hat{\mathcal{G}}.addEdge(i,j,t_{ij})$\;
	}	
\end{algorithm}

From the paths build in Algorithm \ref{alg:1}, the inferred social network $\hat{\mathcal{G}}$ is constructed with the additional edges in Algorithm \ref{alg:2}. Here, for each index vertex pair in $hPaths$, maximum $t_{ij}$ is computed from all the possible heuristic paths in $hPaths[\langle i,j \rangle]$. Now, $\mathcal{P}(len(k))$ is calculated as $(len(k)-1)/L_{max}$ and $\mathcal{B}(\theta)$ is computed as mentioned in the previous seection. Considering all propagation, the number of paths is $\mathcal{O}(nd^{L_{max}})$. However, applying the heuristic, this count becomes $o(nd^{L_{max}})$. As it has to traverse all the paths in $hPaths$, the complexity of the algorithm becomes $o(nd^{L_{max}})$. 

\subsection{Heuristic Covering All Possible Edges} \label{sec:heu_all}
In every recursive call of Algorithim \ref{alg:1} based on the value of $\theta^c$ some of the vertices are not considered for further propagation. Due to this, some of the edges are missed out in $\hat{\mathcal{G}}$, if they are reached from the source only via the left out vertices. Now, to recover all possible edges from enumeration techniques (if the heuristic is not applied), we modify the Algorithm \ref{alg:1} by adding the function \texttt{checkPath()} mentioned in Algorithm \ref{alg:3}. \texttt{checkPath()} is called at \textbf{else} part of the \textbf{if} condition at Line number 12 of the Algorithm \ref{alg:1}. The function \texttt{checkPath()} works in the similar way as the \texttt{addNeighbor()} function does in the Algorithm \ref{alg:1}. However, it only adds an entry in $hPaths$, if it has not been found yet. This process prserves the same density as of its enumeration counterpart with comparitively less time. 

\begin{algorithm}
	\caption{Function for checking path existence in $hPaths$}
	\label{alg:3}
	\SetKwFunction{FMain}{checkPath}
	
	\SetKwProg{Pn}{Function}{:}{\KwRet}
	\Pn{\FMain{$sNode$, $cNode$, $\mathcal{G}$, $L$, $path$}}{
		\If{ $L==1$}{  
			\textbf{return}\;
		}
		\For{$i \in \mathcal{G}.succesors(cNode)$}{
			\If{$i \notin path$ and $! \ \mathcal{G}.hasEdge(sNode,i)$}{ 
				$tempPath = path $\;
				$tempPath.append(i)$\;
				\If{ 	$hPaths[\langle sNode,i \rangle]$ not exist}{  
					$hPaths[\langle sNode,i \rangle].append(tempPath)$\;
				}
				\texttt{checkPath}{($sNode$, $i$, $\mathcal{G}$, $L-1$, $tempPath$)}\;
				
			}	
		} 
	}
\end{algorithm}

\subsection{Recommendation using Inferred $\hat{\mathcal{G}}$} 
In memory based trust-aware recommendation \cite{ massa2007trust, golbeck2005personalizing, pal2017trust, Moradi:2015:RRM:2827896.2828216}, the rating is predicted in similar to the user-based collaborative filtering \cite{Herlocker:1999:AFP:312624.312682}. The trust weight is used in place of user-user similarity score and the weighted average is taken from the user's trusted neighbors in $\hat{\mathcal{G}}$. The predicted rating $\hat{r}_{i,j}$ for an user $u_i \in \mathcal{U}$ to an item $i_j \in \mathcal{I}$ is given in Equation \ref{eq:ratpred}, where $\bar{r_i}$ is the mean rating of user $u_i$.

\begin{equation}\label{eq:ratpred}
\hat{r}_{i,j} = \bar{r}_i + \frac{\sum_{{u_k} \in \Gamma_{\hat{\mathcal{G}}}(u_i)}{t_{i,k} \cdot (r_{k,j} - \bar{r}_k )}}{\sum_{{u_k} \in  \Gamma_{\hat{\mathcal{G}}}(u_i)}{t_{i,k}}}
\end{equation}

\section{Experimental Results and Discussions} \label{sec:results}
In this section, we discuss the experimental setup, datasets description, obtained results and their analysis. All the experiments are carried out on Intel Xeon 40-core processor, 64GB memory server. 

\paragraph{Datasets Used} In our experiment, we use two publicly available and real-world datasets(downloaded from librec data\footnote{https://www.librec.net/datasets.html}). For Epinions, we start with random 5000 users. In both the datasets users are selected, if present in both the rating data and the social network data. The items are selected if rated by at least two users. The statistics of the datasets are given in Table \ref{tb:dataset}.

\begin{table}[h]
	\centering
	\caption{ \fontsize{8}{8}\selectfont{\uppercase{ Dataset Description}}   } \label{tb:dataset}
	\begin{tabular}{|c|c|c|c|c|c|}
		\hline
		Datatest & \#users & \#items & \#rating & \#social tie & density \\
		\hline
		FilmTrust & 507 & 1888 & 14272 & 1448 & 0.0052 \\
		Epinions & 3446 & 14890 & 78017 & 26303 & 0.0022 \\
		\hline
	\end{tabular}
\end{table}

\begin{table*}[h]
	\begin{center}
		\caption{  \fontsize{8}{8}\selectfont{\uppercase{ Results of Different Metrics in the Inferred Network }}  } \label{tb:Results_1}
		\begin{tabular}{|c|c|c|c|c|c|c|c|c|c|} 
			\hline
			Dataset & $L_{max}$ & Metric & All Path & H1-Th-$indeg$ & H1-Th-$\delta$ & H1-Th-$\gamma$ & H2-Th-$indeg$  & H2-Th-$\delta$ & H2-Th-$\gamma$ \\
			\hline
			
			\hline
			\multirow{28}{*}{FilmTrust} & 3&	Duration (sec) & 0.9410 & 0.6700 & 0.7160 & 0.8930 & 1.0000 & 0.9620 & 0.9901 \\
			
			& 3&	Path Count & 102896 & 84961 & 89396 & 80426 & 90472 & 93640 & 87023 \\
			
			& 3&	\#Edges & 32972 & 29403 & 30124 & 28772 & 32972 & 32972 & 32972 \\ 
			
			& 3&	Density & 0.1176 & 0.1049 & 0.1074 & 0.1026 & 0.1176 & 0.1176 & 0.1176 \\ 
			
			& 3&	Edges Missed (\%) & - & 10.8243 & 8.6376 & 12.7381 & 0.0 & 0.0 & 0.0 \\
			
			& 3&	Score (\%) & - & 11.0943 & 8.9075 & 13.1445 & 0.6005 & 0.4276 & 1.7863 \\
			
			& 3&	Mean Error  & - & 0.5398 & 0.5559 & 0.5942 & 0.0719 & 0.0786 & 0.0127 \\

			\cline{2-10}
			& 4&	Duration (sec) & 8.1935 & 5.3363 & 6.2693 & 5.5833 & 6.6694 & 7.6984 & 7.3714 \\
			
			& 4&	Path Count & 929541 & 590491 & 666159 & 515271 & 612606 & 682746 & 537706 \\
			
			& 4&	\#Edges & 60512 & 46749 & 50899 & 48549 & 60512 & 60512 & 60512 \\ 
			
			& 4&	Density & 0.2158 & 0.1667 & 0.1815 & 0.1732 & 0.2158 & 0.2158 & 0.2158 \\ 
			
			& 4&	Edges Missed (\%) & - & 22.7442 & 15.8861 & 19.7696 & 0.0 & 0.0 & 0.0 \\
			
			& 4&	Score (\%) & - & 24.7224 & 18.2971 & 23.2697 & 6.9374 & 4.8222 & 9.7650 \\
			
			& 4&	Mean Error  & - & 0.5103 & 0.4835 & 0.4761 & 0.1081 & 0.0974 & 0.1272 \\

			\cline{2-10}
			&5&	Duration (sec) & 78.7035 & 38.1192 & 47.2987 & 34.3739 & 56.1122 & 62.0365 & 56.4992 \\
			
			&5&	Path Count & 7986967 & 3843896 & 4683858 & 3101015 & 3889438 & 4716654 & 3145202 \\
			
			&5&	\#Edges & 83180 & 58547 & 68621 & 64700 & 83180 & 83180 & 83180 \\ 
			
			&5&	Density & 0.2967 & 0.2088 & 0.2447 & 0.2308 & 0.2967 & 0.2967 & 0.2967 \\ 
			
			&5&	Edges Missed (\%) & - & 29.6141 & 17.5030 & 22.2169 & 0.0 & 0.0 & 0.0 \\
			
			&5&	Score (\%) & - & 34.7030 & 24.4758 & 29.6502 & 19.4542 & 12.5944 & 18.5429 \\
			
			& 5&	Mean Error  & - & 0.4816 & 0.3952 & 0.4174 & 0.1228 & 0.0999 & 0.1580 \\
			
			\cline{2-10}
			& 6&	Duration (sec) & 3296.0235 & 259.1278 & 357.0994 & 214.1842 & 448.5976 & 514.3984 & 392.8445 \\
			
			& 6&	Path Count & 65056116 & 23125078 & 30951589 & 17631853 & 23189934 & 30998835 & 17696655 \\
			
			& 6&	\#Edges & 96884 & 68768 & 82098 & 75846 & 96884 & 96884 & 96884 \\ 
			
			& 6&	Density & 0.3456 & 0.2453 & 0.2928 & 0.2705 & 0.3456 & 0.3456 & 0.3456 \\ 
			
			& 6&	Edges Missed (\%) & - & 29.0203 & 15.2615 & 21.7146 & 0.0 & 0.0 & 0.0 \\
			
			& 6&	Score (\%) & - & 39.8745 & 27.7693 & 33.4802 & 29.8336 & 20.2293 & 26.0941 \\
			
			& 6&	Mean Error  & - & 0.4371 & 0.3223 & 0.3851 & 0.1418 & 0.1069 & 0.1803 \\
			\hline
			
			\hline
			\multirow{14}{*}{Epinions} & 3&	Duration (sec) & 156.9770 & 104.5001 & 117.1252 & 103.2702 & 147.7529 & 148.2081 & 145.6534 \\
			
			& 3&	Path Count & 14712986 & 8699044 & 10810054 & 8984098 & 9764541 & 11480437 & 9949892 \\
			
			& 3&	\#Edges & 2278134 & 1601709 & 1877875 & 1701186 & 2278134 & 2278134 & 2278134 \\ 
			
			& 3&	Density & 0.1919 & 0.1349 & 0.1582 & 0.1433 & 0.1919 & 0.1919 & 0.1919 \\ 
			
			& 3&	Edges Missed (\%) & - & 29.6921 & 17.5696 & 25.3255 & 0.0 & 0.0 & 0.0 \\
			
			& 3&	Score (\%) & - & 33.3883 & 21.3352 & 29.1207 & 6.9182 & 4.4702 & 6.0593 \\
			
			& 3&	Mean Error  & - & 0.4021 & 0.3806 & 0.5092 & 0.0662 & 0.0676 & 0.0316 \\

			\cline{2-10}
			& 4&	Duration (sec) & 4724.0824 & 2445.1803 & 2452.5392 & 1800.3410 & 3388.3601 & 3666.4240 & 3404.5793 \\
			
			& 4&	Path Count & 401023321 & 191137540 & 209378186 & 146794897 & 140910171 & 212757811 & 150953019 \\
			
			& 4&	\#Edges & 5741955 & 3778982 & 4229831 & 3726220 & 5741955 & 5741955 & 5741955 \\ 
			
			& 4&	Density & 0.4837 & 0.3183 & 0.3563 & 0.3139 & 0.4837 & 0.4837 & 0.4837 \\ 
			
			& 4&	Edges Missed (\%) & - & 34.1865 & 26.3346 & 35.1054 & 0.0 & 0.0 & 0.0 \\
			
			& 4&	Score (\%) & - & 46.3414 & 39.9751 & 48.0087 & 29.8498 & 21.1622 & 26.2018 \\
			
			& 4&	Mean Error  & - & 0.3336 & 0.3028 & 0.3850 & 0.1071 & 0.0914 & 0.0846 \\
			
			\hline
		\end{tabular}
	\end{center}
\end{table*}

\paragraph{Experimental Setup}
In all the datasets, we create inferred network $\hat{\mathcal{G}}$ in the following three ways, 1) using all path enumeration techniques (All Path), 2) using the threshold based heuristic (H1), (Algorithm \ref{alg:1}) mentioned in Section \ref{sec:threshold_heuristic}, and 3) using the heuristic covering all edges (H2), (Algorithm \ref{alg:1} + \ref{alg:3}) mentioned in Section \ref{sec:heu_all}. In both the heuristics, we apply the threshold on three different weight selection in-degree ($indeg$), Degree-of-Trustworthiness($\delta$) and  Degree-of-TrustNPurcahse ($\gamma$). The results for the mentioned methods are denoted as H1-Th-$indeg$, H1-Th-$\delta$, likewise. We construct the graphs for different $L_{max}$ starting from length 2, to compare the processing time. For FilmTrust, we take the maximum length as 6. Whereas for Epinions, we compute up to the length 4 due to the time constraint. Now, for $\theta_c$ calculation, $c_{th}$ is chosen as 1 for indegree as weight and $10*L_{max}$ for other two weight functions. For recommendation using $\hat{\mathcal{G}}$, we perform an offline experiment using leave-one-out mechanism and evaluate the performance using MAE, RMSE and Coverage \cite{pal2017trust}. These terms are defined as follows,

\begin{equation*}
MAE = \frac{\sum \vert r_{ij} - \hat{r}_{ij} \vert }{{\mathcal{R}}}  \ \ ,
\end{equation*}

\begin{equation*}
RMSE = \sqrt{\frac{\sum ( r_{ij} - \hat{r}_{ij} )^2 }{{\mathcal{R}}}} \ \ ,
\end{equation*}

\begin{equation*}
Coverage = (1 - \frac{\mathcal{R}^{np}}{\mathcal{R}}) * 100 (\%)  \ \ ,
\end{equation*}

where $\mathcal{R}$ is the total number of ratings and $\mathcal{R}^{np}$ is the total number of ratings that the recommendation algorithm is unable to predict.

%
%
%

%

\begin{figure*}[h]
	\centering
	\begin{subfigure}{0.25\textwidth}
		\centering
		\includegraphics[width=1\linewidth]{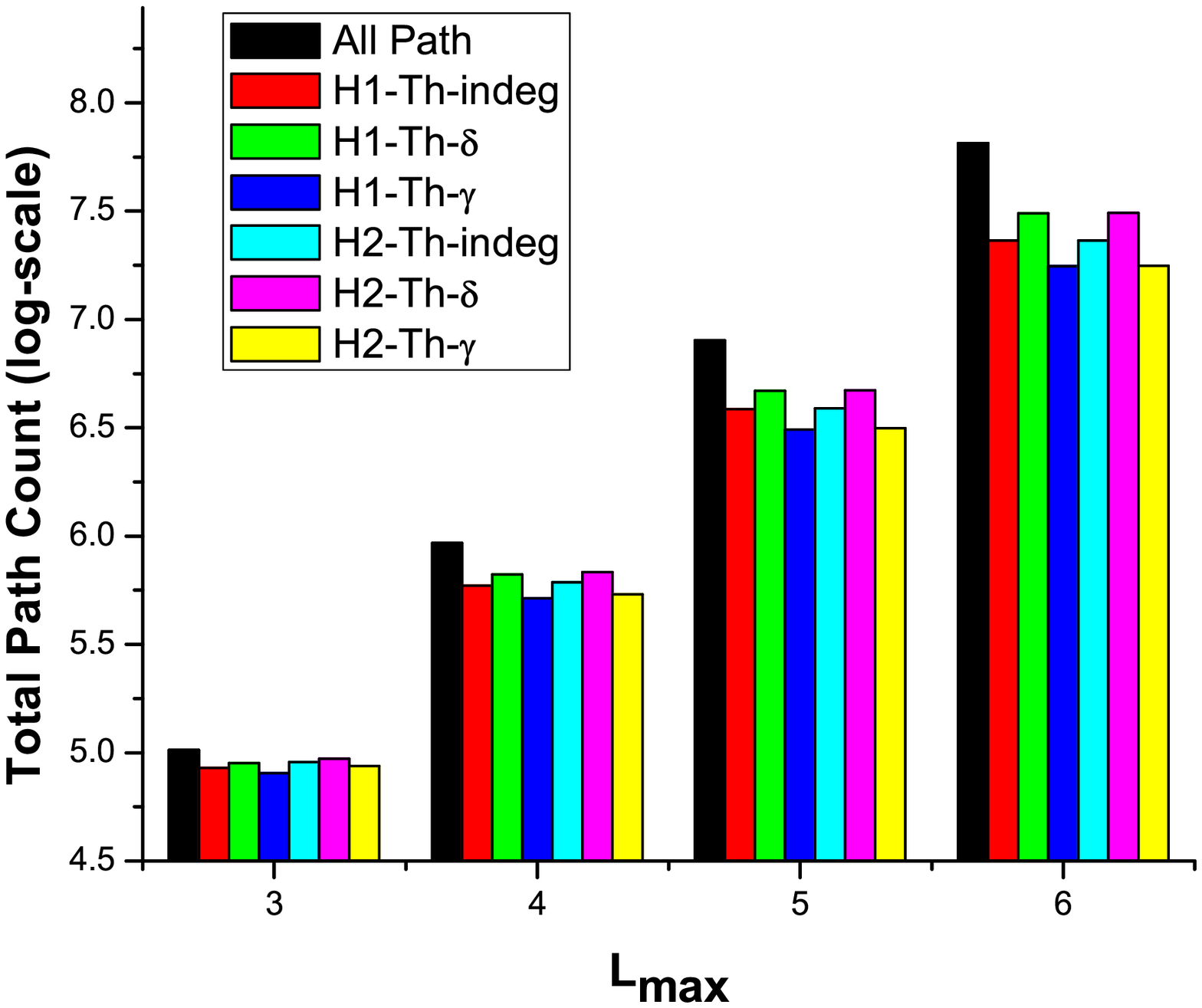}
		\caption{}
	\end{subfigure}
	\begin{subfigure}{0.24\textwidth}
		\centering
		\includegraphics[width=1\linewidth]{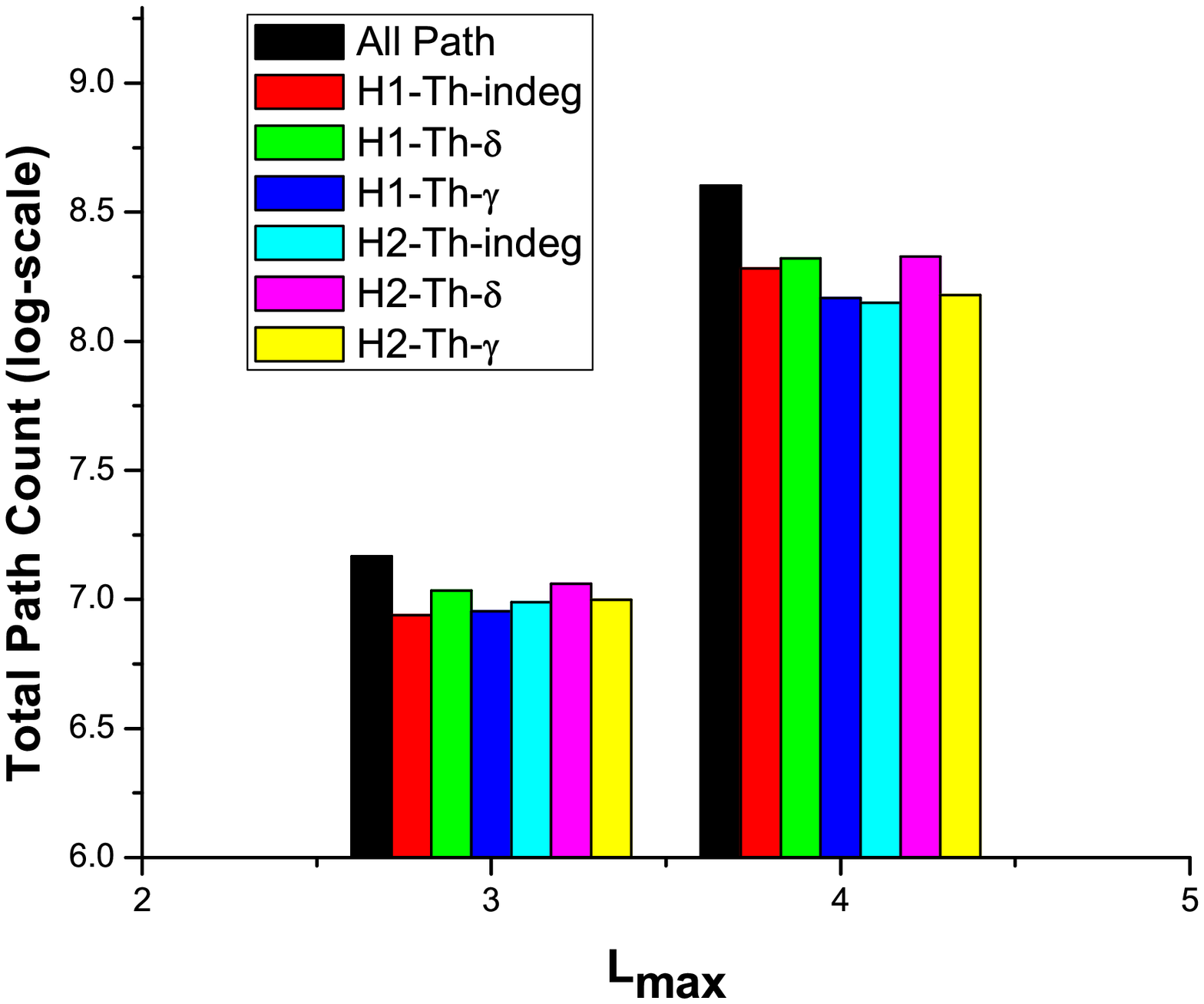}
		\caption{}
	\end{subfigure}
	\begin{subfigure}{0.25\textwidth}
		\centering
		\includegraphics[width=1\linewidth]{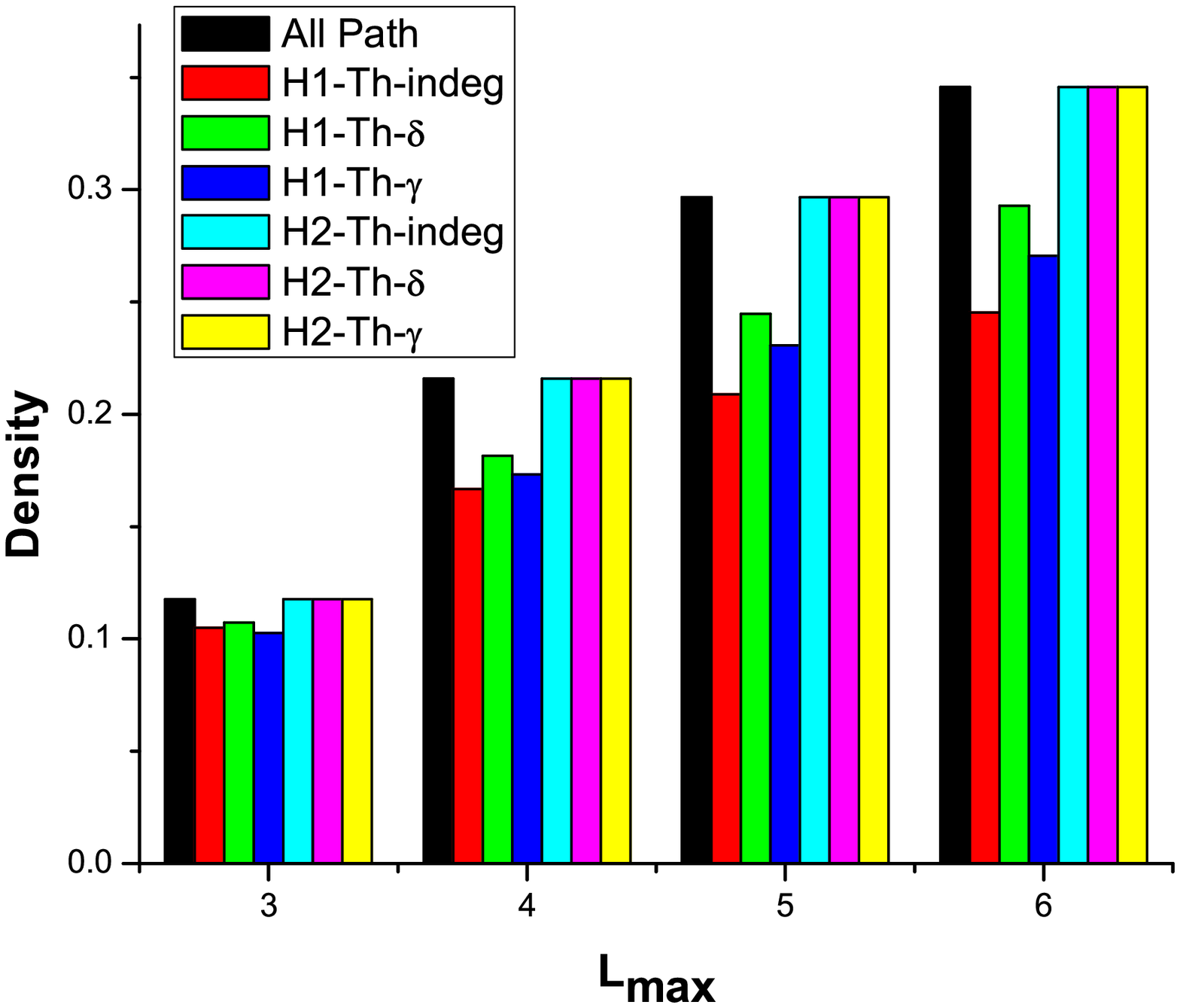}
		\caption{}
	\end{subfigure}
	\begin{subfigure}{0.24\textwidth}
		\centering
		\includegraphics[width=1\linewidth]{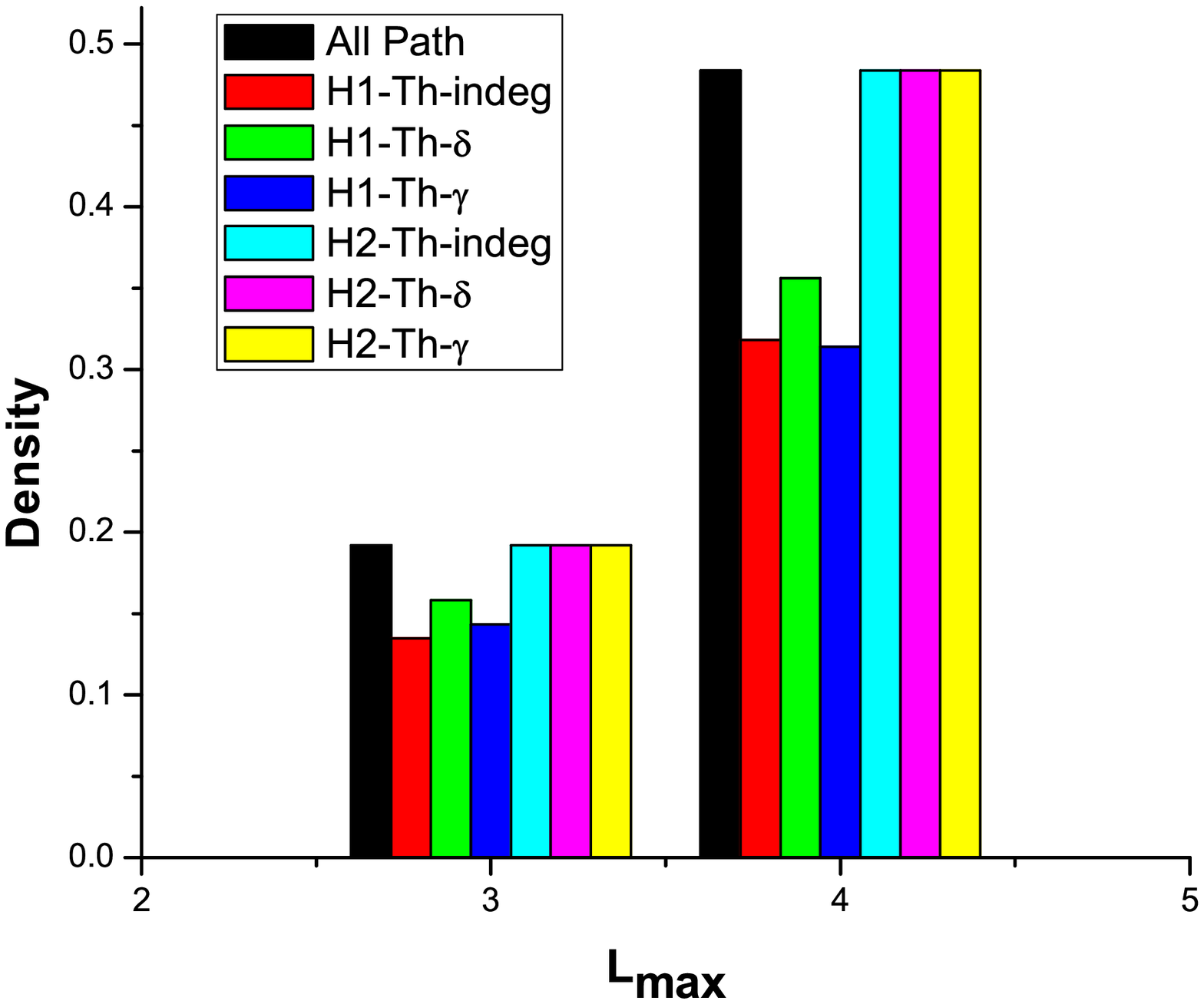}
		\caption{}
	\end{subfigure}
	\caption{ \fontsize{8}{10}\selectfont{(a),(b) show the plots for path count in log scale versus $L_{max}$ for FilmTrust and epinions data; (c),(d) show the plots for density versus $L_{max}$ for FilmTrust and epinions data. } } \label{fig:result}
\end{figure*}

\begin{figure*}[h]
	\centering
	\begin{subfigure}{0.25\textwidth}
		\centering
		\includegraphics[width=1\linewidth]{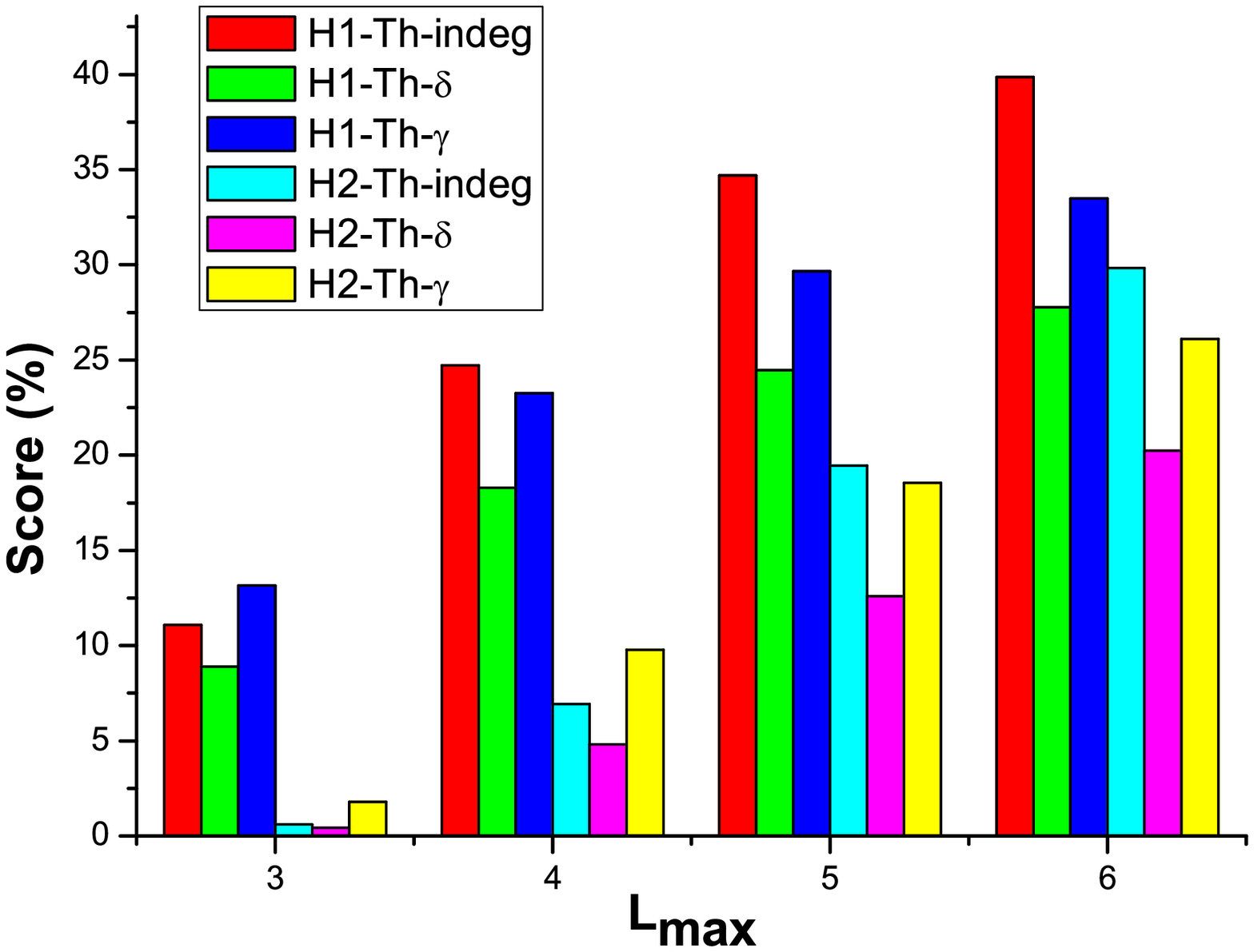}
		\caption{}
	\end{subfigure}
	\begin{subfigure}{0.24\textwidth}
		\centering
		\includegraphics[width=1\linewidth]{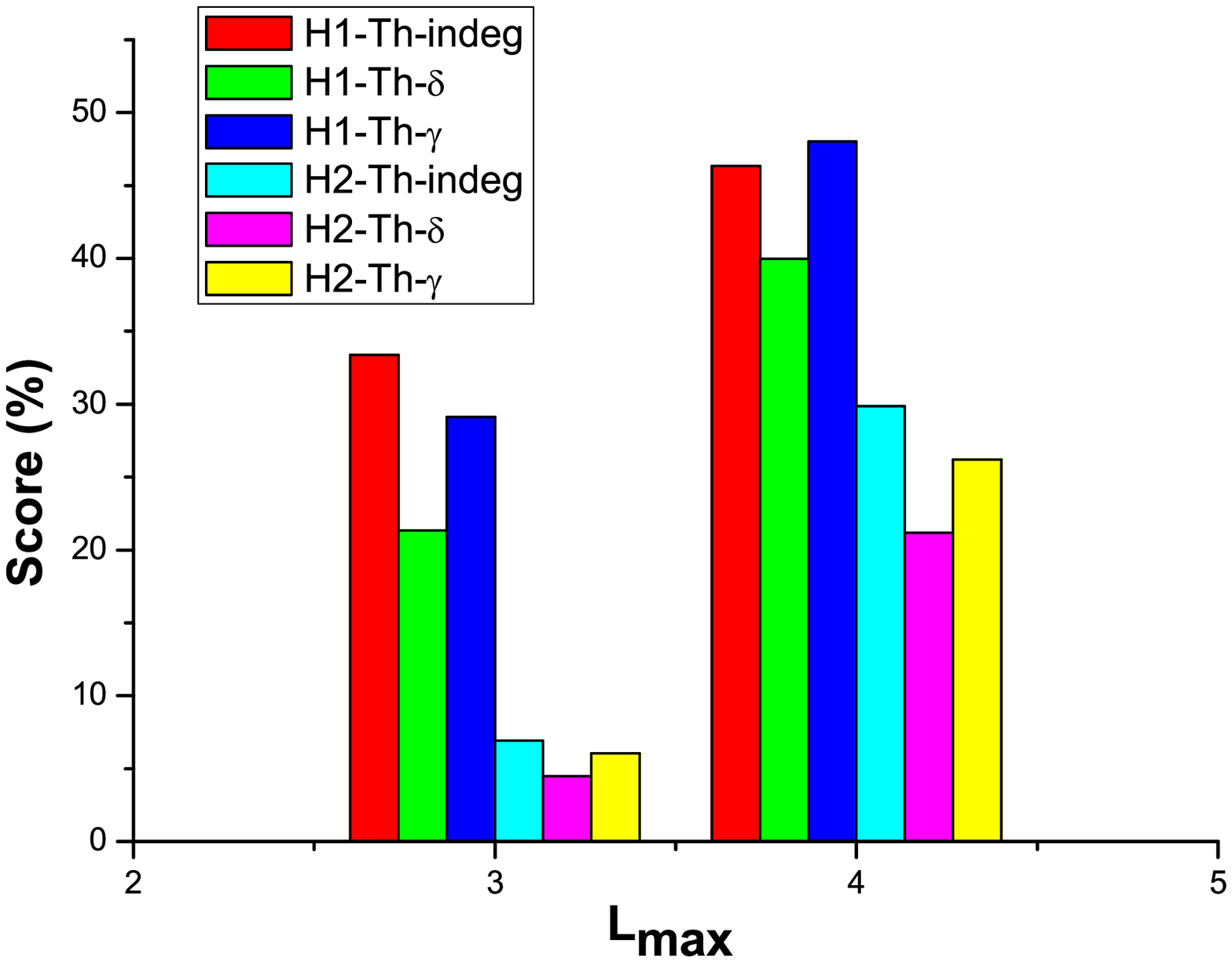}
		\caption{}
	\end{subfigure}
	\begin{subfigure}{0.25\textwidth}
		\centering
		\includegraphics[width=1\linewidth]{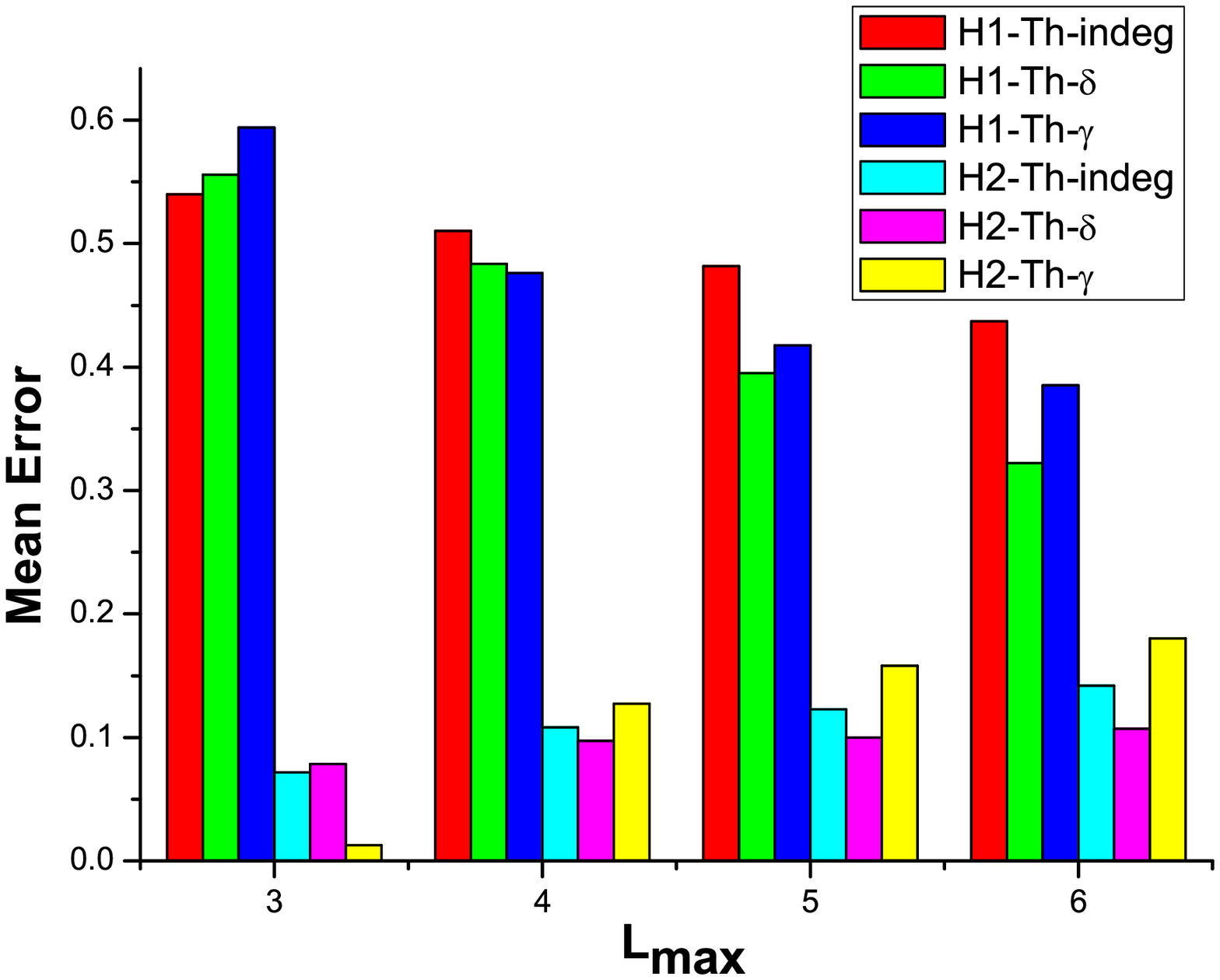}
		\caption{}
	\end{subfigure}
	\begin{subfigure}{0.24\textwidth}
		\centering
		\includegraphics[width=1\linewidth]{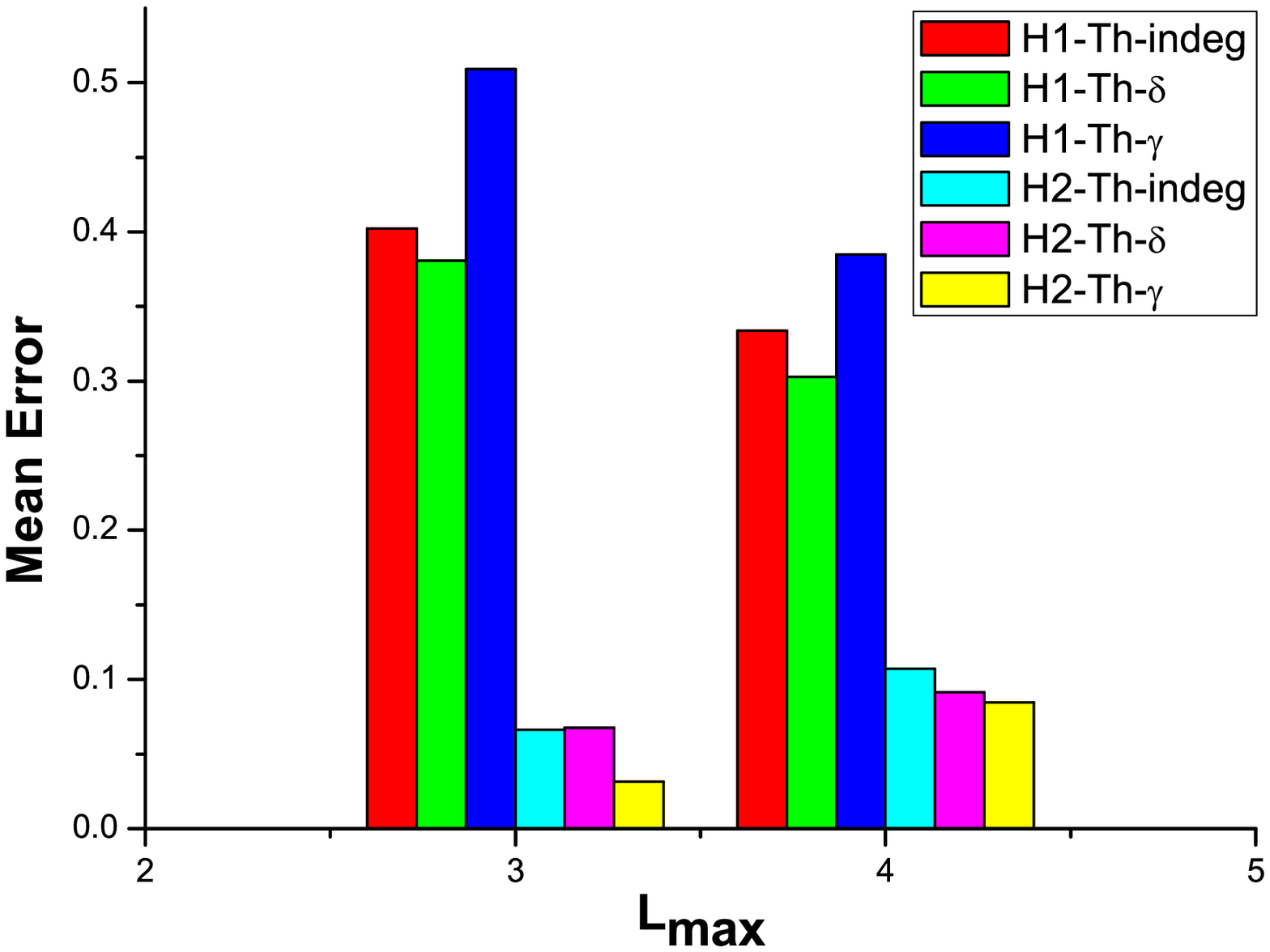}
		\caption{}
	\end{subfigure}
	\caption{ \fontsize{8}{10}\selectfont{(a),(b) show the plots for score(\%) versus $L_{max}$ for FilmTrust and epinions data; (c),(d) show the plots for mean error versus $L_{max}$ for FilmTrust and epinions data.}} \label{fig:result_2}
\end{figure*}

\paragraph{Goals of the Experiment} In the context of different heuristic, the observations are based on the properties of the inferred graphs which have a significant role in the recommendation. We want to study the following facts:

\begin{itemize}
	\item How the duration and the total path counts in $hPaths$ varies for each propagation length.
	\item How the edge-count and density of $\hat{\mathcal{G}}$ changes across different length and heuristic-threshold combination.
	\item Percentage of missing edges in different heuristics from its enumeration counterpart.
	\item The average error in maximum $t_{ij}$ from the heuristic-based techniques.
	\item For a particular vertex weight, how the variation of threshold changes the inferred graph structure.
	\item Observe the changes in recommendation accuracy.
\end{itemize}

\paragraph{Results and Discussion} The results of seven metrics are reported for the comparison of different inferred graphs in Table \ref{tb:Results_1}. For analysis, we discuss them across different length and heuristic, first metric wise, then dataset wise. 

The first two metrics, the duration of computing the paths along with graph creation, and the total path count in $hPaths$, both are dependent on each other. Though the path count is independent of implementation. As the maximum propagation length grows, the path count increases exponentially, which is obvious (shown in Figure \ref{fig:result}). Now, heuristic wise, H1 can discover less number of paths than H2. Also, along with $L_{max}$, the difference of the PathCount in the heuristic base methods and `All Path' method increases. This directly impacts on the duration which also increases exponentially with $L_{max}$.

Next, the number of edges and the density both are dependent on each other. One of them is discussed, and the other one is same. Along with path length, the density increases linearly (shown in Figure \ref{fig:result}). In, all the threshold based heuristics of H1, the density becomes less than `All Path' and H2, where H2 preserves the same density with `All Path'. Now, as the edges are missed in H1, we capture the percentage of missed out edges. We observe that around 30\% edges are missed when the value of $L_{max}$ is 6 for FilmTrust dataset and 4 for Epinions data. This percentage increases along with $L_{max}$. 

Next is another two related metric Score(\%) and Mean Error. Score defines the percentage of edge-count for which suboptimal inferred trust is obtained due to the heuristic based trust propagation. Here, the optimal signifies the maximum trust coming from all possible paths. Mean Error gives the average error in suboptimal trust. As the absence of an edge means 0 trust value, the Score value is higher in H1 than H2. Also, another important point to notice here is that, the Mean Error increases along with $L_{max}$ in H2 but decreases in H1. However, the Score is always increasing along with $L_{max}$ for both H1 and H2 (shown in Figure \ref{fig:result_2}).

As the Epinions dataset is comparatively larger than FilmTrust, the PathCount becomes significantly huge for smaller value of $L_{max}$. So, the proposed heuristics work much faster and gives better approximation of all the possible inferred edges.

\begin{figure}[h]
	\centering
	
	\includegraphics[scale=0.3]{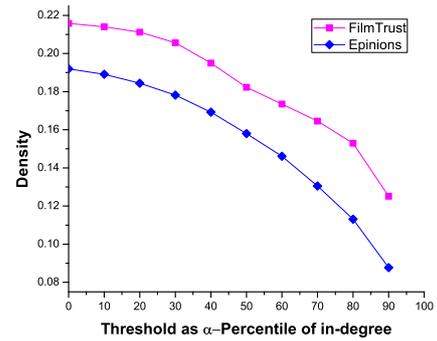}
	\caption{ \fontsize{8}{10}\selectfont{Density Comparison for different $\alpha$ Threshold}} \label{fig:alpha}
\end{figure}

\paragraph{Sensitivity Analysis of Different Threshold Selection} 
Here, we do an experiment by setting the different threshold for subset $\mathcal{X}$ construction. We choose indegree as the vertex weight function. For every propagation from $cNode$, the sorted indegree list is computed from its immediate neighbors' indegree. Then, from the list, the threshold degree is chosen as the $\alpha$ percentile of the degree values. The nodes having indegree greater than the threshold degree is propagated further. We conduct the experiments on FilmTrust and Epinions datasets with length 4 and length 3 inferred graph respectively. Figure \ref{fig:alpha} plots the results on density comparison with $\alpha$ from 10 to 90 percentile. If the value of $\alpha$ is less, then the threshold becomes smaller, which leads to larger subset $\mathcal{X}$ for trust propagation. Now, in comparison with the respective density mentioned in Table \ref{tb:Results_1}, Figure \ref{fig:alpha} shows that H1-Th-indeg reaches similar density at around 70 percentile in both the datasets.

\begin{table*}[h]
	\centering
	\caption{ \fontsize{8}{8}\selectfont{\uppercase{ Results of Recommendation Accuracy }}  } \label{tb:Result_2}
	\begin{tabular}{|c|c|c|c|c|c|c|c|c|c|}
		\hline
		Dataset & Evaluation & All-$\delta$ & All-$\gamma$ & H1-Th-$indeg$ & H1-Th-$\delta$ & H1-Th-$\gamma$ & H2-Th-$indeg$  & H2-Th-$\delta$ & H2-Th-$\gamma$ \\
		\hline
		\hline
		FilmTrust & MAE & 0.6591 &  0.6563 & 0.6562 & 0.6561 & 0.6569 & 0.6564 & 0.6561 & 0.6565\\
		FilmTrust & RMSE & 0.8609 & 0.8610 & 0.8605 & 0.8613 & 0.8605 & 0.8611 & 0.8609 & 0.8610\\
		FilmTrust & Coverage(\%) & 67.42 & 67.42 & 66.41 & 67.18 & 66.70 & 67.42 & 67.42 & 67.42\\
		\hline
		Epinions & MAE & 0.5422 & 0.5421 & 0.5464 & 0.5456 & 0.5459 & 0.5426 & 0.5425 & 0.5421\\
		Epinions & RMSE & 0.7032 & 0.7029 & 0.7089 & 0.7077 & 0.7087 & 0.7039 & 0.7036 & 0.7030\\
		Epinions & Coverage(\%) & 93.48 & 93.48 & 89.47 & 90.55 & 89.09 & 93.48 & 93.48 & 93.48\\
		\hline
	\end{tabular}
\end{table*}

\paragraph{Comparison with Recommendation Accuracy}
Using the inferred social network coming from all the heuristics and `All Path', the rating is predicted, as in TARS. The $L_{max}$  is chosen as the maximum one from the respective datasets. The results of MAE, RMSE and coverage is shown in Table \ref{tb:Result_2}. Coverage determines the percentage of ratings the algorithm is able to predict. If it is unable, then the predicted rating is replaced with the global mean rating. Compared to FilmtTrust, in Epinions dataset, TARS shows better performance in all the evaluation metrics. In FilmTrust dataset, all the methods perform similarly. This signifies that the heuristics are able to maintain the recommendation accuracy. The only difference is noticed in the coverage of H1. However, this small variation in coverage can be neglected. All the H1 methods have higher MAE, RMSE than `All Path' and H2, which is caused due to difference in density. Whereas, H2 is able to maintain almost same prediction accuracy with `All Path'. This signifies that H2 works much accurately than H1 in a large graph with the cost of slightly more time.

%
%


\section{Conclusion} \label{sec:conclusion}
In this paper, we have proposed two threshold based heuristics for trust inference in a social network. We have also presented a cut-off threshold selection strategy on the three proposed vertex weight consideration. We have analyzed different metrics in the inferred graph. The reported results confirm that the heuristic-based methods are able to recover approximately 70\% of the edges from the all possible path selection strategy. We have also used the inferred trust values for recommendation and observed that the heuristic methods are capable of preserving recommendation accuracy. The tighter bound on time complexity of the proposed algorithms can be investigated further. In the heuristic based trust inference, along with the trust values, distrust among users can also be incorporated.

\section*{Acknowledgment}
The work has been financially supported by the project \textit{E-business Center of Excellence} funded by MHRD, Govt. of India under the scheme of \textit{Center for Training and Research in Frontier Areas of Science and Technology (FAST)}, Grant No. F.No.5-5/2014-TS.VII.



%
\bibliographystyle{IEEEtran}
\bibliography{paper.bib}
\end{document}